\documentclass[prl,letterpaper,twocolumn,showpacs]{revtex4} 
\usepackage{times,xspace} 
\usepackage{amsbsy,amssymb,amsmath,bm} 
\usepackage{graphicx,color,epsfig,rotate} 
\usepackage{fancyhdr} 
 
\def\bbbc{{\mathchoice {\setbox0=\hbox{$\displaystyle\rm C$}\hbox{\hbox 
to0pt{\kern0.4\wd0\vrule height0.9\ht0\hss}\box0}} 
{\setbox0=\hbox{$\textstyle\rm C$}\hbox{\hbox 
to0pt{\kern0.4\wd0\vrule height0.9\ht0\hss}\box0}} 
{\setbox0=\hbox{$\scriptstyle\rm C$}\hbox{\hbox 
to0pt{\kern0.4\wd0\vrule height0.9\ht0\hss}\box0}} 
{\setbox0=\hbox{$\scriptscriptstyle\rm C$}\hbox{\hbox 
to0pt{\kern0.4\wd0\vrule height0.9\ht0\hss}\box0}}}}

\newcommand{\ignore}[1]{} 
\newcommand{\mComment}[1]{} 
\newcommand{\gComment}[1]{} 
\newcommand{\jComment}[1]{} 
\newcommand{\rComment}[1]{} 
\newcommand{\lComment}[1]{} 
 
\renewcommand{\mComment}[1]{\textcolor{blue}{Manny: #1}} 
\renewcommand{\gComment}[1]{\textcolor{red}{Gerardo: #1}} 
\renewcommand{\jComment}[1]{\textcolor{green}{Jim: #1}} 
\renewcommand{\rComment}[1]{\textcolor{magenta}{Ray: #1}} 
\renewcommand{\lComment}[1]{\textcolor{purple}{Rolando: #1}} 
 
\begin{document} 
\title{Ground state and thermal transitions in Field Induced spin-Supersolid Phase} 
\author{P. Sengupta$^{1,2}$ and C. D. Batista$^1$} 
\affiliation{$^1$Theoretical Division, Los Alamos National Laboratory, Los Alamos, NM 87545 \\ 
$^2$ MPA-NHMFL, Los Alamos National Laboratory, Los Alamos, NM 87545} 
 
\date{\today}

\begin{abstract} 
We use a quantum Monte Carlo method to study the ground state and thermodynamic
phase transitions of the spin supersolid phase in the $S=1$ Heisenberg
model with uniaxial anisotropy. The thermal melting of the supersolid phase 
shows unqiue signatures in
experimentally measurable observables. This Hamiltonian is a particular case of 
a more general and ubiquitous model that describes the low energy spectrum of a 
class of {\it isotropic} and  {\it frustrated} spin systems. We also discuss 
some alternative realizations of spin supersolid states in real magnets. 
\end{abstract} 
 
 
\maketitle %
\thispagestyle{fancy} 
 The supersolid (SS) state of matter has attracted great interest lately 
following the experiments of Kim and Chan on solid $^4$He. While it is still 
unclear whether it can be stabilized in the continuum, there are several numerical studies
which show that a SS phase can be realized in the presence of a periodic 
potential or underlying lattice for bosons \cite{Sengupta05,Boninsegni05,Triangular}
as well as spins\cite{Ng06,Sengupta07}. The SS state is easier to stabilize 
on a lattice  because the lattice parameter of the ``solid phase" or charge density 
wave cannot relax to any arbitrary value (it has to be an integer multiple of the 
underlying lattice parameter). In this work we have studied a class of spin-SS on 
cubic lattices, focusing primarily on the unique signatures of the thermal
melting of the SS in experimentally measurable observables. We also discuss
different conditions under which a spin-SS can be realized in real spin compounds.
 
The minimal spin model that has a thermodynamically stable supersolid ground state 
on a bipartite lattice is the $S=1$ Heisenberg model with uniaxial 
single--ion and exchange anisotropies and an external magnetic field: 
\begin{equation} 
H_H = J \sum_{\langle {\bf i,j} \rangle } (S^x_{\bf i} S^x_{\bf j} + S^y_{\bf i} S^y_{\bf j}  
+\Delta S^z_{\bf i} S^z_{\bf j} ) + \! \sum_{\bf i} (D {S^z_{\bf i}}^2 - B S^z_{\bf i}) 
\label{eq:H} 
\end{equation} 
where $\langle {\bf i,j} \rangle$ indicates that ${\bf i}$ and ${\bf j}$ are nearest  
neighbor sites, $D$ is the amplitude of the single ion-anisotropy and $\Delta$  
determines the magnitude of the exchange uniaxial anisotropy.  
Note that although the exchange interaction is anisotropic, the longitudinal ($J$) and  
transverse ($\Delta$) couplings are both AFM (positive). 
Henceforth, $J$ is set to unity and all the parameters are expressed in units of $J$. 
 
The ground state properties of the above model on a square lattice were studied in detail
previously\cite{Sengupta07}. For $D,\Delta > 1$, the ground state is supersolid  over a 
finite range of applied field $B$. In this work, we report the ground state and
thermodynamic properties of (\ref{eq:H}) on a cubic lattice. While the quantum
phase diagram remains qualitatively unchanged, the thermodynamic
properties are different in three dimensions (3Ds) because the condensate (XY
ordered antiferromagnetic phase) extends to 
finite temperatures. Additionally, the melting of this phase belongs to the XY 
universality class as opposed to Kosterlitz-Thouless (KT) type in two dimensions (2Ds). 
This has important consequences in any putative experimental detection of the SS phase, 
as we shall discuss below.   

The  Stochastic Series expansion (SSE) quantum Monte Carlo (QMC) method\cite{Sandvik} is used to
study the Hamiltonian (\ref{eq:H}) on cubic lattices $N=L\times L\times L/2$, with
$8\leq L \leq 16$. To characterize the different phases, we compute the longitudinal 
component of the staggered static structure factor (SSSF), 
\begin{equation} 
 S^{zz}({\bf Q})={1\over N}\sum_{j,k}e^{-i{\bf q}\cdot({\bf r}_j-{\bf r}_k)}\\  
\langle  S^z_jS^z_k\rangle, \hskip0.2in {\bf Q}=(\pi,\pi,\pi),
\end{equation} 
and the spin stiffness, $\rho_s$, defined as the response of the system to a 
twist in the boundary conditions. $S^{zz}({\bf Q})$ measures the extent of
diagonal (Ising like) long-range order (LRO) at the ordering wave vector ${\bf Q}=(\pi,\pi,\pi)$,
while the stiffness (superfluid density in particle language \cite{Batista04}), 
indicates the presence of XY (off-diagonal)  LRO (this is not true for D$<3$). 
In 3D, the superfulid density is identical to the 
condensate fraction. In the simulations, the stiffness is obtained from  
the winding numbers of the world lines along the three
axes: $\rho_s=\langle W_x^2+W_y^2+W_z^2\rangle/3\beta$\cite{Ceperley}. 

{\it Ground state (GS) phases} As the field, $B$, is varied, the GS of (\ref{eq:H})
goes through a succession of phases, including spin-gapped Ising-ordered (IS)  
and gapless $XY$-ordered ($XY$) states. The IS phase is marked by a diverging value 
of $S^{zz}({\bf Q}) \propto N $  in the thermodynamic limit, whereas
a finite $\rho_s$ characterizes the gapless XY ordered phase. 
A spin SS phase is identified by a finite value of both 
$S^{zz}({\bf Q})/N $ {\em and\/} $\rho_s$ \cite{note0}. Both 
quantities are always finite for finite size systems and estimates for $N \to \infty$ 
are obtained from finite-size scaling.  
 
\begin{figure}[] 
\includegraphics[angle=0,width=9cm]{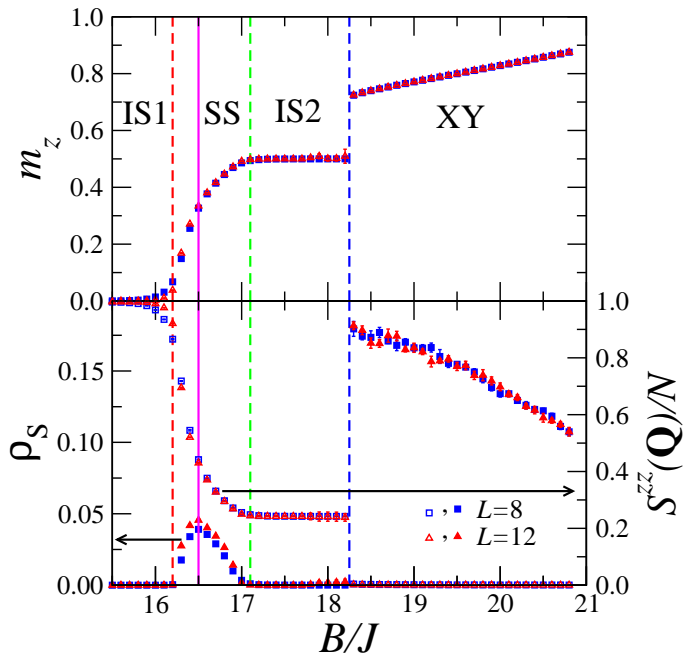} 
\caption{(Color online) Quantum phase diagram of $H_H$   
(Eq.~\ref{eq:H}) for $D=3.0$ and  $\Delta=6.0$. The upper panel
shows the magnetization 
as a function of field $B$. The SS phase appears between the  
two Ising-like phases IS1 and IS2. At higher fields, 
there is a first order transition to a pure XY--AFM phase.  
The lower panel shows the stiffness and the longitudinal component of
the SSF. The SS phase has finite values of both observables.}
\label{fig:gs} 
\end{figure}

Fig.~\ref{fig:gs} shows the quantum phase diagram as a function of magnetic field, 
$B$, for $D=3.0$ and $\Delta=6.0$ -- it is qualitatively similar
to that obtained in 2Ds\cite{Sengupta07}. The $m_z (B)$ curve features two prominent 
plateaus corresponding to different IS phases. For small $B$, the GS 
is a gapped AFM solid (IS1)  with no net magnetization. The stiffness, $\rho_s$, 
vanishes in the thermodynamic limit, while $S^{zz}({\bf Q})/N \approx 1$  
with the spins primarily in the $S^{z}_{{\bf i}}=\pm 1$ states in the two sublattices.
At a critical field, $B_{c1}$, there is a second order transition to a state with a 
finite fraction of spins  
in the $S^z_{{\bf i}}=0$ state. These $S^z=0$ ``particles'' Bose-Einstein condense (BEC)
to give the GS a finite stiffness. The diagonal order is reduced but remains finite
as well. The resulting GS thus has simultaneous long-range diagonal
and off-diagonal order; in other words, it is a spin-supersolid.
The complete phase diagram consists of a second gapped Ising phase (IS2) with
diagonal order (all the spins in the $S^z=-1$ sublattice are flipped to $S^z=0$) 
and an $XY$ phase at very high fields with pure off-diagonal ordering.
 
{\it Finite temperature transitions} Finite temperature properties of the 
SS has previously been examined for hard core bosons\cite{Schmid04,Boninsegni05,
Kawashima07} and $S={1\over 2}$ spins on a bilayer\cite{Laflorencie07}. The
melting of the SS phase proceeds via two steps -- the superfluid order 
disappears at a lower temperatute whereas the solid order persists up to a 
higher temperature. In 2Ds, the continuos U(1) symmetry cannot be broken at 
$T>0$ and the SS has only a quasi long-range off-diagonal order for
$T<T_{KT}$. The vanishing of the spin stiffness occurs via a KT transition.
In contrast, true long-range off-diagonal order in the SS persists to
finite temperatures in 3Ds and the melting of the superfluid order belongs to the
XY universality class. The solid order disappears at a higher temperature 
via an Ising-like transition.

\begin{figure}[] 
\includegraphics[angle=0,width=9cm]{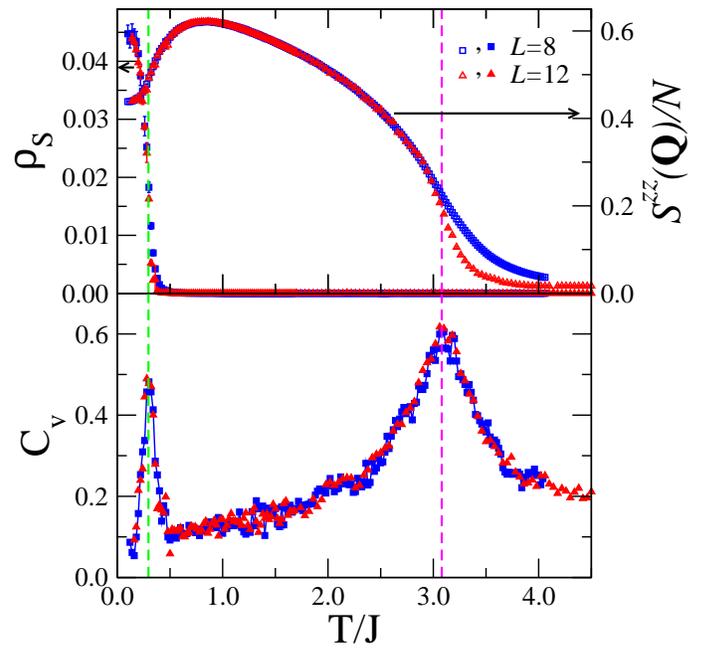} 
\caption{(Color online) Two-step melting of the SS for parameters in Fig.1
and $B/J$=16.5 (solid line in Fig.1). The top panel shows the vanishing
of the two orders at different temperatures. The disappearance of 
superfluidity is accompanied by an unusual increase in the solid order parameter. The 
lower panel shows the signatures in the specific heat
at the two transitions.} 
\label{fig:ssmelting} 
\end{figure}

The results of simulations of thermal transitions associated with the SS
phase are shown in fig.\ref{fig:ssmelting}. The top panel shows the variation 
of the solid and superfluid order parameters as a function of temperature. At
low temperatures, both order parameters are finite. With increasing $T$, the
SS ``melts'' into a pure solid. The disappearence of SF order is marked by
an enhancement in the solid order. This apparently anomalous behavior reflects 
the fact that in the SS phase, the
solid order is partially suppressed by the co-existing SF order.  The
longitudinal component of the SSF is accessible experimentally by 
neutron scattering and its non-monotonic behavior at the onset of superfluid order can serve as an
important signature of the SS phase.
The three dimensionality of the model implies that the two transitions 
should be accompanied by specific heat anomalies at the corresponding 
temperatures. The XY transition will manifest itself as a 
$\lambda$-anomaly while the Ising-like solid melting will be marked by a cusp.
Indeed we find clear signatures of the two transitions in the calculated
specific heat (lower panel of fig.\ref{fig:ssmelting}). While both the
peaks are rounded by finite-size effects, their positions coincide
unambiguously with the melting of the superfluid and Ising orders. 
Since it is one of the most readily measurable observables,
having clear signatures in the specific heat is of great
experimental relevance. 
 
\begin{figure}[!htb] 
\hspace*{-1.1cm} 
\includegraphics[angle=90,width=9cm]{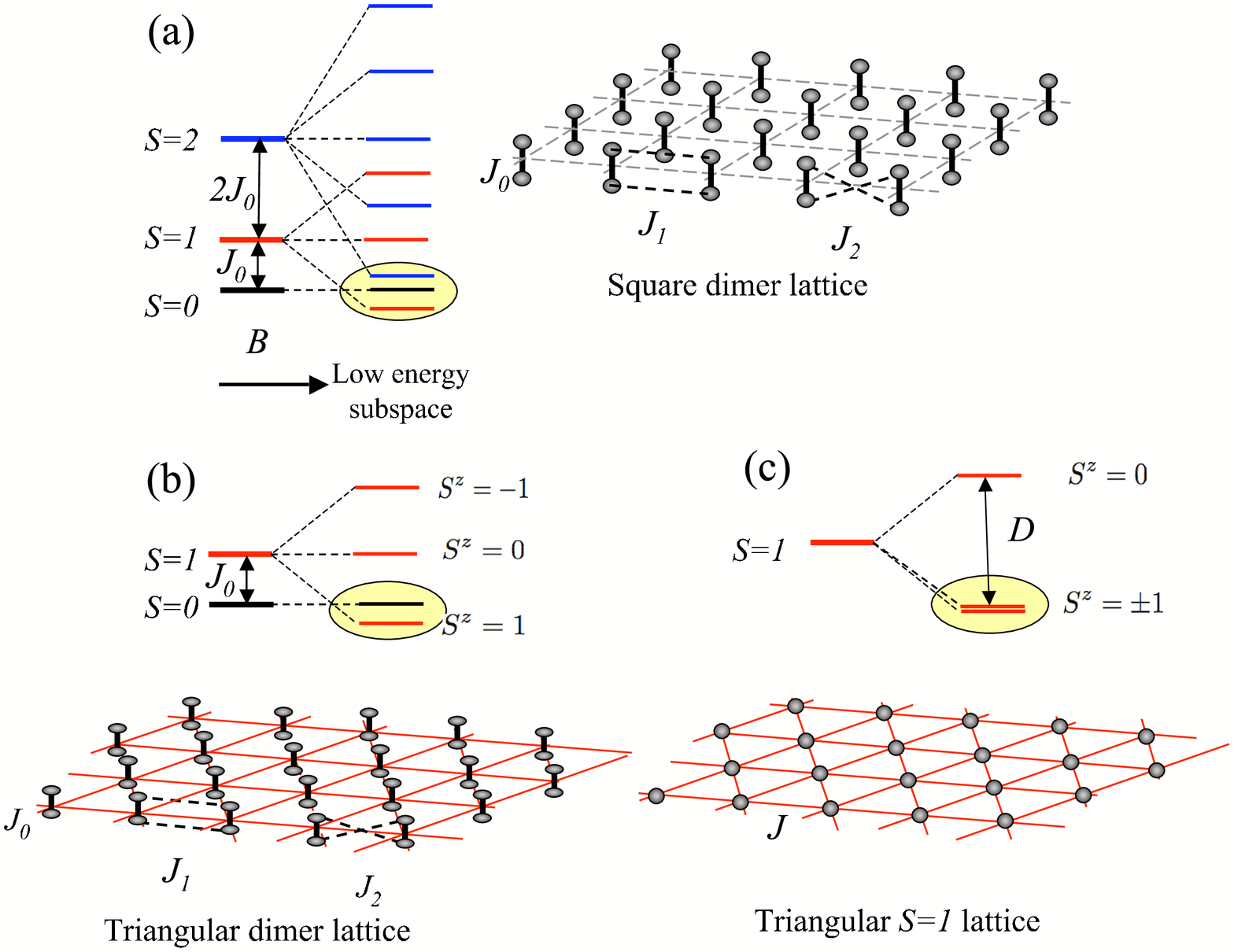}
\caption{(Color online) (a) Square lattice of S=1 dimers with an intra-dimer Heisenberg 
AFM interaction $J_0$ and inter-dimer interactions $J_1$ and $J_2$. The level diagram  
shows the low energy subspace of the single dimer spectrum in the presence of a 
magnetic field. (b) $S=1/2$ dimers on a triangular lattice and the low energy subspace 
of a single dimer. (c) $S=1$ spins on a triangular lattice and the energy level
splitting for easy-axis single-ion anisotropy.} 
\label{dimer} 
\end{figure}

Next we discuss the relevance of these results for finding a SS phase in  
real magnets. The magnetic properties of spin compunds with spin-orbit 
interaction  much smaller than  the crystal field splitting can be 
adequately described by a U(1) invariant model (although
this invariance is never perfect)\cite{note0}. The transition metal magnetic  
ions belong to this class. On the other hand,  the exchange 
anisotropy is typically very small in these compunds. The above model with large $\Delta$
is not directly applicable to this class of real quantum magnets. We shall
show below that under appropriate conditions, an effective uniaxial exchange
anisotropy can be generated in the low-energy subspace of a model with
(realistic) isotropic interactions. To this end we consider coupled layers
of dimers with
only {\it isotropic} (Heisenberg) AFM interactions -- an intra-dimer 
exchange $J_0$ and weaker inter-dimer {\it frustrated} couplings $J_1$ and $J_2$
(see Figs.~\ref{dimer}(a) and \ref{dimer}(b)):  
\begin{eqnarray} 
H_D &=& J _0\sum_{\bf i} {\bf S}_{\bf i+} \cdot {\bf S}_{\bf i-} + 
J_1 \sum_{\langle {\bf i,j} \rangle, \alpha}  {\bf S}_{\bf i\alpha} \cdot {\bf S}_{\bf j\alpha}\nonumber \\  
&+&J_2 \sum_{\langle {\bf i,j} \rangle, \alpha}{\bf S}_{\bf i\alpha} \cdot {\bf S}_{\bf j{\bar \alpha}} - B \sum_{{\bf i}\alpha} S^z_{\bf i \alpha}. 
\label{eq:hdimer}
\end{eqnarray} 
The index $\alpha=\pm$ denotes the two spins on each dimer. 

For $S=1$ dimers, the 
low energy subspace of $H_D$ (for $J_1,J_2 \ll J_0$) consists of the singlet, 
the $S^z=1$ triplet and the $S^z=2$ quintuplet (see Fig.~\ref{dimer}(a)). 
The low energy  effective model that results from restricting $H_D$ to this 
subspace supports a field-induced supersolid phase on a bipartite lattice 
for $J_0 > z(J_1+J_2)/2$ and $J_0 \gg z(J_1-J_2)/2$ ($z$ is the co-ordination 
number of the lattice) as was shown in Ref.\cite{Sengupta07}.

For $S=1/2$ dimers the low energy
subspace of $H_D$ consists of the singlet and the $S^z=1$ triplet states in the limit 
$J_0 \ll J_1,J_2$ (see Fig.~\ref{dimer}(b)). The resulting low--energy effective
 model is a $t-V$ Hamiltonian for hard core bosons:
\begin{equation}
H_{eff}=-t\sum_{\langle {\bf i,j} \rangle}(b_{\bf i}^{\dagger}b_{\bf j} + b_{\bf j}^{\dagger}b_{\bf i}) + V\sum_{\langle {\bf i,j} \rangle}n_{\bf i}n_{\bf j} - \mu\sum_{\bf i}n_{\bf i}
\label{eq:heff}
\end{equation}
$b_{\bf i}^{\dagger}$ creates a $S^z=1$ triplet state at site {\bf i} whereas 
the singlet corresponds to the empty boson state; $n_{\bf i}$ is the boson
number operator $b_{\bf i}^{\dagger}b_{\bf i}$ and the parameters of the 
effective model are expressed in terms of those of the original Hamiltonian 
$H_D$ as $t=(J_1-J_2)/2$, $V=(J_1+J_2)/2$ and 
$\mu= -J_0 + B$. On many frustrated lattices, this
model contains a SS phase in its quantum phase diagram for $t<0$ and $V\gg |t|$
\cite{Boninsegni05,Triangular,Kawashima07}. In terms of the original model, this 
implies that $S=1/2$ dimers with frustrated inter--dimer couplings, $J_2\gtrsim J_1$
 provides an alternative realization of a spin-SS on different frustrated lattices.

As a final example, we consider $S=1$ Heisenberg
model with large easy-plane single-ion anisotropy ($\Delta=1$ and $D<0$ in Eq.(1)). 
For $|D| \gg J$ the low-energy subspace consists of the $S_i^z=\pm 1$ states 
(see Fig.~\ref{dimer}(c)). The low--energy effective model is once again
the $t-V$ Hamiltonian (\ref{eq:heff}) with $t = -J^2/2D$, 
$V=J+J^2/D$ and $\mu=B-J^2/D-2n_bJ$, up to second order in $J/D$. $n_b$ is the number of
bonds per site. As in the previous case, a SS phase is realized for $V\gg |t|$ 
on different frustrated lattices\cite{Boninsegni05,Triangular,Kawashima07}.
The BEC component corresponds to spin nematic ordering.

In conclusion, we have used numerical simulation to study the ground state
and thermodynamic phase transitions involving the spin-supersolid phase in
a $S=1$ Heisenberg model with uniaxial exchange and single-ion anisotropies on
a cubic lattice. The melting of the SS occurs in two steps with the $XY$ and
Ising ordering disappearing at different temperatures. The transitions are
marked by unique features in the structure factor and the specific heat which
will be useful in any experimental detection of the SS. Finally, we
discuss several different conditions under which a SS can be realized in real 
spin compunds. 

LANL is supported by US DOE under Contract No. W-7405-ENG-36.


\begin{references}
\bibitem{note0}
In real systems, spin--orbit interactions will remove the finite stiffness 
by opening a gap in the 
excitation spectrum. However, the U(1)--invariant model considered
here provides a very good description of the static properties
of real systems as long as the temperature is higher than the 
the small U(1)--symmetry breaking terms.
\end{references}
\end{document}